\documentstyle[preprint,floats,tighten,prl,aps,epsf]{revtex}
\def \pom {{\scriptscriptstyle \kern -0.1em I \kern -0.25em P}}
\def\desepsf(#1 width #2){\epsfxsize=#2 \epsfbox{#1}}
\begin{document}
\preprint{\vbox{
\hbox{PSU-TH/216} 
\hbox{August 1999}
\hbox{Revised November 1999}
\hbox{} }  }
\draft

\title{Infrared divergences and non-lightlike eikonal lines \\ 
in Sudakov processes}
\author{J.C.\ Collins and F.\ Hautmann}
\address{Department of Physics, Pennsylvania State University, 
University Park PA 16802}

\maketitle

\begin{abstract}%
We propose a method to analyze infrared contributions to 
non-inclusive processes in QCD. 
We use the one-loop Sudakov form factor as a working example. 
Borrowing techniques from renormalization theory, we construct 
counterterms for the contributions from the soft and 
collinear regions, and  we relate them  to eikonal Wilson
lines taken along non-lightlike directions. 
\end{abstract} 

\pacs{}

%\narrowtext

\section{Introduction} 

Understanding the structure of hadronic final states in high energy
collisions involves applying QCD in situations with multiple mass
scales.  While the application of QCD to inclusive hard-scattering
processes characterized by a single large mass scale is well under
control, the use of QCD in less inclusive, multiple-scale processes is
much subtler~\cite{tasi95}.  

A common feature of these more general situations is that soft gluon
effects do not cancel.  In tractable cases they merely factorize, and
result in a more general factorization theorem, with new kinds of
sensitivity to non-perturbative physics.  An example is the transverse
momentum distribution for the Drell-Yan process~\cite{sumform}.  But
we would like to treat, for example, Monte-Carlo event generators,
where the theory does not yet \cite{monte}
go systematically beyond the leading
logarithm approximation, despite the great practical importance of
these event generators.  The necessary extensions of the factorization
theorems require that we devise new more precise methods.  This is
the aim of the present paper.

We consider the well-known example of the one-loop part of the Sudakov
form factor, Fig.~\ref{figformfac}a. A full factorization theorem was
obtained in \cite{jccmue}.  Our method is a systematic subtraction
method that generalizes the Bogoliubov $R$-operation for
renormalization. (For a related approach see Ref.~\cite{tka}.)
 The subtractions are obtained from matrix
elements of certain gauge-invariant operators, which is essential if
the factorization theorem is to be considered a real theorem of field
theory.  Moreover the subtractions are performed on the integrand, so
that our final result for the ultraviolet part of the graph is
infrared safe: it can be numerically integrated.  
Although our analysis is at the one loop level, the method should be
generalizable to higher orders.

The plan of the paper is as follows. In Sec.~\ref{IR.subs} we describe
the method and the form of the counterterms, which are obtained from
matrix elements of path-ordered exponentials of the gluon field along
non-lightlike lines.  In Sec.~\ref{evolution} we discuss the evolution
due to changes in these directions. 
  In Sec.~\ref{outlook} we summarize and briefly discuss
results and developments.

\section{The infrared subtraction method} 
\label{IR.subs}

\begin{figure}
\centerline{\desepsf(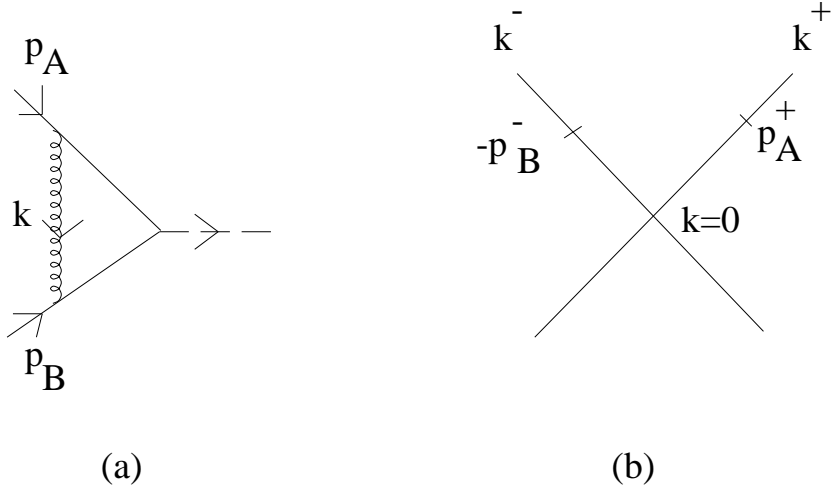 width 11 cm)}
\vspace*{5mm}
\caption{(a) Form factor graph; (b) soft and collinear regions. }
\label{figformfac}
\end{figure}

Consider the graph in Fig.~\ref{figformfac}a.
To simplify the calculations while 
retaining all the essential ingredients on which  we want to focus,
we work in a massive abelian theory with scalar quarks.
We denote the quark mass by $m$ and the gauge boson mass by $m_g$. 
We  work in a center-of-mass frame in which the incoming quark momenta  
$p_A$ and  $p_B$ are  in the $+z$ and $-z$ directions, with  
$ 2 \,  p_A^+ \, p_B^- = Q^2$. 
We will compute the asymptotic large $Q$ behavior of the graph when
the integrand is multiplied by a test function $\varphi(k/Q)$:
\begin{eqnarray}
\label{vertex}
\Gamma [ \varphi ] &=& i g^2 \int {{d^4 k} \over {(2 \, \pi)^4}}  
{{ (2 \, p_A - k) \cdot (2 \, p_B + k) } 
\over {  \left( k^2 - m_g^2 + i \, \varepsilon \right) \, 
\left[ (p_A - k)^2 - m^2 + i \, \varepsilon \right] \, 
\left[ (p_B + k)^2 - m^2 + i \, \varepsilon \right] \,  } } \, 
\varphi ({ k / Q}) 
\nonumber\\
  && - ~\mbox{$\overline{\rm MS}$ counterterm for UV divergence}
\hspace*{0.1 cm} . 
\end{eqnarray} 
For the actual graph of  Fig.~\ref{figformfac}a we will have 
$\varphi = 1$. But retaining $\varphi({ k / Q})$ until the end of 
the calculation will assist us in identifying the nature of the terms 
we obtain.  It will also enable us to immediately solve a more general
class of problem, when an infrared finite vertex correction is
inserted. 

Following the analysis of~\cite{jccmue,libby,korchrady} 
we  identify 
the important regions of momentum space in $\Gamma$,  
as depicted in Fig.~\ref{figformfac}b: 
\begin{enumerate}
\item soft region: all components of $k^\mu$ are much smaller than $Q$: 
$k^\mu \sim \lambda \, Q$, where $\lambda$ is small;  

\item $p_A$-collinear region: $k^+ \sim Q$, $k_\perp \sim \lambda \, Q$, 
$k^- \sim \lambda^2 \, Q$, with 
$ 0 \leq k^+ \leq p_A^+$;  

\item $p_B$-collinear region: $k^- \sim Q$, $k_\perp \sim \lambda \, Q$, 
$k^+ \sim \lambda^2 \, Q$, with 
$ - p_B^- \leq k^- \leq 0$;  

\item hard region: all components of $k^\mu$ are of order $Q$: 
$k^\mu \sim  Q$. 
\end{enumerate}
We will construct a term for each region. We will do this starting from 
the smallest region
(region 1, representing a  
0-dimensional subspace) and systematically  going to larger regions:
regions 2 and 3 (1-dimensional) and region 4 (4-dimensional). 
The strategy for constructing the 
approximation for a given region will be   
\begin{itemize}
\item[a)] take the original graph and subtract the terms for smaller 
regions; 

\item[b)] construct an approximation valid in the region being treated; 

\item[c)] show that it has no divergences or leading contributions 
from smaller regions; 

\item[d)] ``renormalize'' divergences associated with larger regions. 

\end{itemize}
Since, as we will see, only regions that are larger or smaller 
than the current region are relevant, the sum of all the terms 
will give a good leading-power approximation to the original graph.

\subsection{Soft region }

We begin with the soft region. Expanding the integrand of $\Gamma$ in
powers of the small parameter $\lambda$ (in the definition of the region)
gives
\begin{equation} 
\label{S0term}
S_0 =  { {-i \, g^2 } \over {(2 \, \pi)^4}} \, 
\int \, d k^+ \, d k^- \, d^2 {\bf k} \, { 1 \over  
{ ( k^2 - m_g^2 + i \, \varepsilon ) }} 
\,  { 1 \over { \left( k^- - i \, \varepsilon \right) \, 
\left( k^+ + i \, \varepsilon \right)   } } \, \varphi ( 0 )
\hspace*{0.3 cm} .     
\end{equation}  
The expansion in $\lambda$ gives the correct answer in the soft region
unless one or both of the components $k^+$, $k^-$ is too small. In
that case $ k^+k^- \ll {\bf k}^2$, and only the quark
propagators have significant dependence on $k^+$ and $k^-$.  For the
original graph, we deform the $k^+$ and $k^-$ contours out of the
dangerous region.  The approximation is only valid on the deformed
contour, so that we must apply the same $i\varepsilon$'s to the approximation
in Eq.\ (\ref{S0term}). Observe that the
direction of the deformation of the contour integrations is
determined by the fact that the interactions are initial-state
interactions.  Therefore the same deformation applies in
all higher-order graphs as well.

Now we observe that Eq.~(\ref{S0term}) is the one-loop contribution to
the vacuum expectation value of two Wilson
lines~\cite{jccmue,korchrady} taken along light-like directions ${\hat
p}_A$, ${\hat p}_B$:
\begin{equation}
\label{S0tilde}
{\widetilde S}_0 = { { 
\langle 0 | T \left[ 
V_q ( {\hat p}_A ) \, V_{\bar q} ( {\hat p}_B ) \right] | 0 \rangle  
 } 
\over { 
\langle 0 | V_q ( {\hat p}_A )  | 0 \rangle \,
\langle 0 |  V_{\bar q} ( {\hat p}_B ) | 0 \rangle }} 
\hspace*{0.2 cm} , 
\end{equation}
where self-energy contributions have been divided out 
and we have defined 
\begin{equation}
\label{VqVqbar}
V_q ( n ) = {\cal P}\exp\left(
 i g \int_{-\infty}^0 dz \, A (z \, n ) 
 \cdot {n} \right)
\hspace*{0.2 cm} ,  \hspace*{0.3 cm} 
V_{\bar q} ( n ) = {\cal P}\exp\left(
 - i g \int_{-\infty}^0 dz \, A (z \, n ) 
 \cdot {n} \right)
\hspace*{0.2 cm} .      
\end{equation}  

Our soft term $S_0$ still has singularities from the collinear and
ultraviolet regions. The ultraviolet singularity is to be dealt with
by the standard subtractive renormalization procedure, by adding a
counterterm.  We will now treat the collinear singularities in a
similar fashion.  The collinear singularities are entirely associated
with the integration over the rapidity $y = \ln ( k^+ / k^- )$ and are
not removed by dimensional regularization~\cite{coltka}.

As usual, the definition of the counterterms is non-unique.  The
reasons for our choice of the definition of the collinear counterterms
are as follows. First, the new term is to have an operator
interpretation, so that we can give it a nonperturbative
definition. Secondly, the evolution equation is to be ``simple'', in a
sense to be specified later.  Here we just observe that the new
definition will involve considering Wilson lines along an arbitrary
direction, and the evolution equation treats the dependence on this
direction. 

Let us introduce two non-lightlike vectors $u_A$, $u_B$:
$u_A = \left( u_A^+, u_A^-, {\bf 0} \right)$, 
$u_B = \left( u_B^+, u_B^-, {\bf 0} \right)$.
Then we define the subtracted soft term by
\begin{eqnarray} 
\label{Sterm}
S &=&  { {-i \, g^2 } \over {(2 \, \pi)^4}} \, 
\int \, d k^+ \, d k^- \, d^2 {\bf k} \, { 1 \over  
{ ( k^2 - m_g^2 + i \, \varepsilon ) }} 
\, \left[ { 1 \over { ( k^- - i \, \varepsilon ) \, 
( k^+ + i \, \varepsilon ) \, } } \right. 
\nonumber\\ 
&&  \hspace*{0.8 cm} 
- \left. 
{ 1 \over { ( k^- - i \, \varepsilon )  } } \, 
{ { u_B^- } \over { (u_B^- \, k^+ + u_B^+ \, k^- + i \, \varepsilon )}} - 
{ { u_A^+ } \over { ( u_A^+ \, k^- + u_A^- \, k^+ - i \, \varepsilon )}} 
\, { 1 \over { ( k^+ + i \, \varepsilon )  } } 
\right] \, \varphi ( 0 ) 
\nonumber\\ 
&& 
- {\mbox {UV}} \hspace*{0.3 cm} .     
\end{eqnarray}  
The first term in the square brackets is the unsubtracted soft term.
The second term subtracts the divergence from the region collinear to
$p_A$, i.e., from large positive rapidity $y$.  The $u_B$ factor in
this term provides a cut-off when the rapidity of $k$ goes much below
the rapidity of $u_B$. Similarly, the third term subtracts the
divergence from the region collinear to $p_B$.  Observe that, in order
to avoid invalidating the contour deformation argument, we must have
$u_A^+, u_B^- > 0$, $u_A^-, u_B^+ < 0$.  Note that the
collinear-to-$p_A$ subtraction term has no collinear-to-$p_B$
singularity; indeed it is power suppressed in this region. The same is
true with $A$ and $B$ exchanged.  Provided that we choose $u_A$ and
$u_B$ to have large rapidities: $u_A^+/u_A^- \gtrsim p_A^+/p_A^-$, and
$u_B^+/u_B^- \lesssim p_B^+/p_B^-$, then $S$ continues to provide a good
approximation to $\Gamma$ in the soft region.  The ultraviolet subtraction
in the third line of Eq.~(\ref{Sterm}) is understood to be done in the
$\overline{\rm MS}$ scheme.

The cut-offs are gauge-invariant since, as we now show, they are
obtained from a Wilson-line operators.  Using the path-ordered
exponentials introduced in Eq.~(\ref{VqVqbar}), we define
\begin{equation}
\label{Spath}
{\widetilde S} = 
\frac{ 
   \langle 0 | V_q ( {\hat p}_A ) \, V_{\bar q} ( {\hat p}_B ) | 0 \rangle \, 
   \langle 0 | V_q ( {u}_A )  | 0 \rangle \, 
   \langle 0 | V_{\bar q} ( {u}_B )  | 0 \rangle 
}{ 
   \langle 0 | V_q ( {\hat p}_A ) \, V_{\bar q} ( {u}_B ) | 0 \rangle \,
   \langle 0 | V_q ( {u}_A ) \, V_{\bar q} ( {\hat p}_B ) | 0 \rangle 
}
\hspace*{0.2 cm} ,
\end{equation}
whose one-loop term gives Eq.~(\ref{Sterm}).  The first factor in the
numerator gives the original soft term $S_0$ in Eq.\ (\ref{S0term}),
and the two factors in the denominator, involving eikonal lines along
spacelike directions, give the collinear subtractions.  The remaining
factors in the numerator cancel factors of a complete external
propagator for the Wilson line.  Given our one-loop formulae, this
result appears to us to be unique, if we simply assume that the
quantity which we calculate is the product of vacuum expectation
values of some Wilson line operators.

\subsection{Collinear to $A$ region} 

Next consider the collinear-to-$p_A$ region.  We will construct a term $A$
such that $A+S$ gives a good approximation to $\Gamma$ (valid to the
leading power) over the whole region, including its soft subregion.
It is obtained by expanding $\Gamma-S$ in powers of the parameter $\lambda$
in the definition of the collinear-to-$p_A$ region, for which $k^+ \sim
Q$, $k_\perp \sim \lambda \, Q$, $k^- \sim \lambda^2 \, Q$.  Thus
\begin{eqnarray} 
\label{Aterm}
A &=&  \left( \Gamma - S \right)_{|p_A - {\rm {coll.}} } 
\nonumber\\
&=&
{ {i \, g^2 } \over {(2 \, \pi)^4}} \, 
\int \, d k^+ \, d k^- \, d^2 {\bf k} \, { 1 \over 
{ \left( k^2 - m_g^2 + i \, \varepsilon \right) }} 
\nonumber\\
&&  \times \left[ 
{ { (2 \, p_A^+ - k^+ ) \, 2 \, p_B^- } \over 
{ \left( - 2 \, p_A^+ \, k^- + 2 \, k^+ \, 
[ k^- - m^2 / (2 \, p_A^+) ]  
- {\bf k}^2  
+ i \, \varepsilon \right) \, \left( 2 \, p_B^- \, k^+ 
+ i \, \varepsilon \right) }} \right. \, \varphi (k^+ /Q,  0 , {\bf 0})
\nonumber\\ 
&&  \left. 
\hspace*{0.8 cm} + { 1 \over { \left( k^- - i \, \varepsilon \right) \, 
\left( k^+ + i \, \varepsilon \right) \, } } \, \varphi (  0 ) - 
{ 1 \over { \left( k^- - i \, \varepsilon \right)  } } \, 
{ { u_B^- } \over { ( u_B^- \, k^+ + u_B^+ \, k^- + i \, \varepsilon )}} 
\, \varphi (  0 ) \right] 
\nonumber\\ 
&& 
- {\mbox {UV}} \hspace*{0.3 cm} .    
\end{eqnarray}
By this construction of $A$, we ensure that $A+S$ is a good
approximation to the collinear-to-$p_A$ region.  As to the soft region,
observe that the first term in square brackets, the collinear
expansion of the original graph, has a soft singularity at $k^+=0$.
But the second term, obtained from $S$, cancels this singularity.
Notice that the third term is power suppressed in the soft region.
Hence $A+S$ is a good approximation to the original graph in a
neighborhood of both the soft and collinear-to-$p_A$ regions.

We now show that the $A$ term, Eq.\ (\ref{Aterm}), is power suppressed
in region 3, where $k$ is collinear to $p_B$.  This region merely
overlaps with the region that this term is designed to treat.  In the
$p_B$-collinear region, a good approximation is obtained by replacing
the quark factor $(2p_A^+-k^+)/[(p_A-k)^2-m^2+i\varepsilon)]$ by the 
light-like eikonal $2p_A^+/(-2p_A^+k^+-i\varepsilon) = -1/(k^++i\varepsilon)$.  
This is the same replacement as for the soft region, so that we get a 
suppression. This is a result and solves what is effectively a problem in
overlapping divergences: We obtained $A$ by expanding about the line
$k^- = {\bf k}_T = 0$ --- Fig.~\ref{figformfac}b, and it is suppressed
at the intersection, $k=0$, with the collinear-to-$p_B$ line.  The
extrapolation back to the whole of the $B$ region continues to be
suppressed.

We therefore have the following results:
\begin{itemize}

\item The sum $A+S$ gives a good approximation to $\Gamma$ in regions 1 and 2.

\item The term $A$ is power suppressed in regions that overlap with
    the $A$ region or are smaller than it.

\end{itemize}

\subsection{Collinear to $B$ region} 

The expression for the $p_B$-collinear term, $B$, is simply obtained
from Eq.~(\ref{Aterm}) by interchanging the indices $+$ and $-$, $A$
and $B$.

Just as for the soft term $S$, the collinear terms can be obtained
uniquely as the one-loop expansion of a product of gauge-invariant
matrix elements: 
\begin{equation}
\label{Apath}
{\widetilde A} = { { 
\langle 0 | V_{\bar q} ( {\hat p}_B ) \, q (0) | p_A \rangle \, 
\langle 0 | V_q ( {\hat p}_A ) \, V_{\bar q} ( {u}_B )  | 0 \rangle } 
\over { 
\langle 0 | V_q ( {\hat p}_A ) \, V_{\bar q} ( {\hat p}_B ) | 0 \rangle \,
\langle 0 | V_{\bar q} ( {u}_B ) | 0 \rangle }} 
% \hspace*{0.2 cm} , 
%\end{equation}
%\begin{equation}
%\label{Bpath}
\hspace*{0.1 cm} , \hspace*{0.2 cm} 
{\widetilde B} = { { 
\langle 0 | {\bar q} (0) \, V_{q} ( {\hat p}_A )  | p_B \rangle \, 
\langle 0 | V_q ( {u}_A ) \, V_{\bar q} ( {\hat p}_B )  | 0 \rangle } 
\over { 
\langle 0 | V_q ( {\hat p}_A ) \, V_{\bar q} ( {\hat p}_B ) | 0 \rangle \,
\langle 0 | V_{q} ( {u}_A ) | 0 \rangle }} 
 \hspace*{0.2 cm} . 
\end{equation}

\subsection{Ultraviolet region } 

Finally, we consider the hard part of the form factor.  We first
observe that the sum of the terms for the smaller regions, $A+B+S$, 
provides a good approximation to $\Gamma$ in the whole of the smaller
regions 1, 2, and 3. To obtain a term
$H$ for the hard region (4), we simply take the massless approximation 
to $\Gamma-A-B-S$
\begin{eqnarray}
\label{Hterm} 
H &=& \left( \Gamma - S - A - B \right)_{| \, m , m_g \to 0}  - {\mbox {UV}} 
\nonumber\\
&=& { {i \, g^2 } \over {(2 \, \pi)^4}} \, 
\int \, d k^+ \, d k^- \, d^2 {\bf k} \, 
{ 1 \over 
{ \left( 2 \, k^+ \, k^- - {\bf k}^2 + i \, \varepsilon \right) \,
 } } 
\nonumber\\
&& \times \, 
\left[ 
{ { 4 \, p_A^+ \, p_B^- + 2 \, p_A^+ \, k^-  - 2 \, p_B^- \, k^+ 
- 2 \, k^+ \, k^- + {\bf k}^2 } \over 
{ \left( - 2 \, p_A^+ \, k^- + 2 \, k^+ \, k^- - {\bf k}^2 
+ i \, \varepsilon \right) \, \left( 2 \, p_B^- \, k^+ 
+ 2 \, k^+ \, k^- - {\bf k}^2 
+ i \, \varepsilon \right) }} \right. \, \varphi ({k / Q}) 
\nonumber\\ 
&&   
- { 1 \over { \left( k^- - i \, \varepsilon \right) \, 
\left( k^+ + i \, \varepsilon \right) \, } } \, \varphi (  0 ) - 
{ { 2 \, p_A^+ - k^+   } \over 
{ \left( - 2 p_A^+ \, k^- + 2 k^+ \, k^- - {\bf k}^2  
+ i \, \varepsilon \right) \, \left(  k^+ 
+ i \, \varepsilon \right) }}  \, \varphi (k^+ /Q ,  0 , {\bf 0}) 
\nonumber\\ 
&&
+ \left. 
{ {2 \, p_B^- + k^-} 
\over 
{ \left(  k^- 
- i \, \varepsilon \right) \,  
\left( 2 \, p_B^- \, k^+  + 2 \, k^+ \, k^- - {\bf k}^2 
+ i \, \varepsilon \right) \,  }}  
\, \varphi (0 ,  k^- /Q, {\bf 0})  \right]  - {\mbox {UV}} 
\hspace*{0.3 cm}   .  
\end{eqnarray} 
Note that the all non-lightlike eikonal terms have canceled.

To verify that the subtractions are correct, we perform the $k^+$,
$k^-$ integrations in the case $\varphi = 1$.  This gives
\begin{eqnarray} 
\label{Hbis}
H &=&    { {-g^2 } \over {8 \, \pi^2}}  
\, \int \,{{ d {\bf k}^2 } \over { {\bf k}^2 }} 
   \left\{ \ln \left( \frac{ {\bf k}^2 }{ Q^2 } \right)
          +i\pi
          \,+\, 
             \frac{1 - {\bf k}^2 / Q^2} {R}
             \left[ \ln \left( \frac{ 1 + R }{ 1 - R } \right) 
                    - i \pi 
             \right]
   \right\} 
 ~-~ {\mbox {UV}} 
\hspace*{0.3 cm} ,        
\end{eqnarray} 
where  $Q^2 =  2 \,  p_A^+ \, p_B^-$ and 
\begin{equation} 
\label{rootdet}
   R = \left\{
       \begin{array}{ll} 
           \sqrt{1- 4 \, {{\bf k}^2 /Q^2}} 
           &
           ~ ~ ~\mbox {if $4 {\bf k}^2 /Q^2 \leq 1$}
           \hspace*{0.3 cm} ,        
       \\[2mm]
           i \, \sqrt{ 4 \, {{\bf k}^2 /Q^2} - 1} 
           &
           ~ ~ ~ \mbox {if $4 {\bf k}^2 /Q^2 \geq 1$}
           \hspace*{0.3 cm} .      
       \end{array}
       \right.
\end{equation}
Observe that, as a result of the infrared subtractions, the 
${\bf k}^2$ integration in Eq.~(\ref{Hbis}) is regular at small 
${\bf k}^2$.  The large ${\bf k}^2$ behavior, on the other hand, has
to be dealt with via an ultraviolet subtraction.

Eqs.~(\ref{Sterm}), (\ref{Aterm}) and (\ref{Hterm}) provide the main
result of the application of our method.  The analysis of
Refs.~\cite{jccmue,libby,korchrady} establishes the list of leading
regions.  For each region, we have defined a term that, added to the
terms for smaller regions, gives a good leading-power approximation to
the original graph in that region. For each term, we have shown that
it does not receive leading contributions from regions that are
smaller or have an overlap with the region being considered.  Thus the
sum of all the terms gives a good leading-power approximation to the
original graph:
\begin{equation} 
\label{sumofall} 
   H + A + B + S = \Gamma + \mbox{nonleading power}
 \hspace*{0.3 cm} . 
\end{equation}

\section{Evolution}
\label{evolution}

The spacelike vectors $u_A$, $u_B$ in the soft and collinear
subtractions are not physical parameters, and the sum $H+A+B+S$ is
independent of them.  The utility of these vectors comes from the fact
that evolution equations in $y_A = \frac12 \ln | u_A^+ / u_A^- |$ and
$y_B = \frac12 \ln | u_B^+ / u_B^- |$ can be applied to the factors in
the factorization formula to extract effects associated with large
logarithms~\cite{sumform,jccmue}.  For example, differentiating
Eq.~(\ref{Sterm}) with respect to $y_A$ gives
\begin{equation} 
\label{evyA}
{{\partial S } \over { \partial y_A}} 
 =  { { i \, g^2 } \over {(2 \, \pi)^4}} \, 
\int \,  { d^4 k \over  
{ \left( k^2 - m_g^2 + i \, \varepsilon \right) }} \, 
{ { u_A^2 } \over { (u_A^+ \, k^- + u_A^- \, k^+ - i \, \varepsilon )^2}}  
 \hspace*{0.1 cm} . 
\end{equation} 
This is to be contrasted with the evolution equations obtained in
\cite{jccmue}, which suffer from power-law corrections.  Our new
method has all the power law corrections in the initial construction
of the $A$, $B$, $S$ and $H$ terms.  Since no further approximations
are needed in solving the evolution equations~\cite{tka}, this can make for
easier and cleaner proofs in more complicated situations.  It also
makes it easier to locate the sources of the power-law corrections:
they are all associated with the approximations made to derive the
factorization.

\section{Outlook} 
\label{outlook}

In this paper we have proposed a systematic subtraction procedure to
analyze infrared contributions to non-inclusive scattering processes
in QCD. The counterterms arise from gauge-invariant Wilson line
operators which are in spacelike directions. The form of these
operators allows simple evolution equations to be obtained, which have
no power-law corrections; this makes it much easier to pinpoint where
the power-law errors in factorization theorems occur.  Observe that, 
even in our one-loop example, the pattern of infrared contributions in
the collinear regions is equivalent to that of overlapping divergences
in renormalization theory.

Uniqueness of the operator formulae follows from the assumption of a
factorized form for the complete amplitudes.  Given the general and
systematic nature of our approach, we expect to generalize it to
all higher orders and to other, more physically interesting, processes.

\vskip 0.4 cm 

This research is supported in part by the US Department of
Energy under grant No.~DE-FG02-90ER-40577.

\end{document}